\documentstyle[12pt,fleqn]{article}

\newcommand{\scrig}{\cal G }

\newcommand{\mpi}{M_{\pi}}

\begin{document}
\begin{center}
{\large
 {\bf Pionic Content of $\rho NN$ and $\rho N\Delta$ Vertex Functions}}
\vspace{0.5in}

 { {\bf Q. Haider} \\
  Physics Department, Fordham University, Bronx, NY 10458 }

 { {\bf L.\ C.\ Liu}\\
  Theoretical Division, T-2, Los Alamos National Laboratory \\
  Los Alamos, NM 87545 }
\end{center}
\begin{abstract}
The dynamical content of $\rho NN$ and $\rho N\Delta$ vertex functions
is studied with a mesonic model. A set of coupled integral equations
satisfied by these vertex functions were solved self-consistently.
These solutions indicate that the dominant mesonic content arises from
di-pion dynamics. With the experimentally determined pion-baryon-baryon
coupling constants and ranges as input, the model predicts a $g_{\rho NN}$
that agrees with the meson-exchange-potential results. On the other hand,
it predicts a smaller $f_{\rho N\Delta}$ and much softer form factors.
Implications of the findings on the use of phenomenological
coupling constants in nuclear reaction studies are discussed.

\end{abstract}
\vspace{1.0in}
PACS index:  14.20.-c,Gk; 21.30.+y; 25.80.-e

\pagebreak

Meson-exchange theories have been very successful in describing
nuclear phenomena in the region of low and intermediate momentum
transfers, where practical methods for solving non-perturbative QCD
are yet to be fully developed. Within the framework of the meson-exchange
theory, understanding the dynamics associated with the $\rho NN$ and
$\rho N\Delta$ vertices is a subject of great interest. It is well known that
$\rho$-exchange gives rise to a nucleon-nucleon ($NN$) tensor force which
has a sign opposite to that arising from the exchange of $\pi$. The partial
cancellation between these tensor forces makes it indispensable to include
$\rho$ in the hadronic description of nuclei and nuclear matter. The
important role of $\rho$ in nuclear structure has further motivated the
study of its role in nuclear reactions. Studying the $\rho$ is, however,
complicated by the fact that, unlike $\pi N\rightarrow \Delta$,
the $\rho N\rightarrow N$ and $\rho N\rightarrow \Delta$ processes do not
occur in free space. Consequently, information on these reactions
were mainly obtained from fitting $NN$ phase shifts with the use of
meson-exchange potentials (MEP). It is, therefore, quite likely that
the MEP parameters may have to be modified for studying processes
other than $NN$ scattering. Indeed, analyses of deep inelastic
scattering data indicate that the range of a pion monopole form factor
cannot exceed 650 MeV.\cite{Fran}\cite{Kuma} This upper limit is nearly
a factor of two smaller than the MEP ranges. Smaller ranges have also
been predicted as a necessary outcome of the boson nature of the
pion.\cite{Schu} A recent study of the $p(p,n)\Delta^{++}$ reaction has
shown that the inclusion of $\rho$-exchange mechanisms actually worsens
the fit.\cite{Jain} This disagreement raises the interesting question as
to what are the appropriate $\rho$-meson parameters to be used in meson
production. These developments have motivated us to carry out a detailed
analysis of the $\rho NN$ and $\rho N\Delta$ vertices by means of a
dynamical model that does not treat the rho-baryon-baryon ($\rho BB'$)
processes as contact interactions. The main goal of our study is to
identify important dynamical contents of the phenomenological coupling
constants.

Our model for the $\rho NN$ vertex is illustrated in fig.1, where the wavy,
dashed, thin, and thick lines denote, respectively, the $\rho$, $\pi$, $N$,
and $\Delta$. The initial and final nucleons are labelled $a$ and $c$, while
the intermediate baryons are labelled $b$ and $b'$. Figure 1 corresponds
to the equation
\begin{equation}
 F_{\rho a;c}= F_{\rho;\pi\pi}G^{(+)}_{\pi_{1}}G^{(+)}_{\pi_{2}}
F_{\pi_{1}a;b}G^{(+)}_{b}F_{\pi_{2}b;c}\
 + F_{a;\pi b}G^{(+)}_{b}F_{\rho b;b'}G^{(+)}_{b'}
F_{\pi b';c}\equiv B_{\rho a;c} + F'_{\rho a;c}\ ,
\label{eq:1}
\end{equation}
where $F_{\rho a;c}$ denotes the $\rho a\rightarrow c$ vertex function.
A similar notation applies to the other vertex functions, with the indices
$b,b'$ representing either an $N$ or a $\Delta$. In Eq.(\ref{eq:1})
$B_{\rho a;c}$ denotes the Born amplitude given by the two triangle diagrams
in fig.1, while $F'_{\rho a;c}$ corresponds to pion corrections of the
$\rho b\rightarrow b'$ vertices, as shown by the four $\pi$-loop diagrams.
Since the solutions of the model, $F_{\rho i;j}$ ($i,j=a,b,b'c$), appear
on both sides of Eq.(\ref{eq:1}), we must solve a set of four coupled
equations of the form of Eq.(\ref{eq:1}) with the baryon indices $ac$
equal to $NN$, $N\Delta,$ $\Delta N,$ and $\Delta\Delta$, respectively.
For succinctness, only the equation corresponding to $ac=NN$ is illustrated
in fig.1. We have solved the coupled equations by iteration, and a
convergence was obtained after five iterations. In principle, one should
also include the $\rho$-loop corrections of the vertices. Because
$M_{\rho}>>M_{\pi}$, these corrections will be much less important
than the $\pi$-loop corrections which, as we shall see, are quite small.
They were thus neglected in our calculations.

It is worth noting that diagrams similar to ours have recently been
considered in ref.\cite{Gari}. We emphasize, however, that there
exist important differences between the analysis in ref.\cite{Gari}
and ours. In ref.\cite{Gari} the leading contributor was assumed
to be the phenomenological three-branch $\rho NN$ vertex while the
triangle and loop diagrams were considered as corrections and treated
perturbatively. Moreover, phenomenological MEP $\rho BB'$ coupling
constants were used as inputs and no coupled equations were solved.
As such, it is equivalent to evaluating only first-order corrections
to the MEP $\rho NN$ coupling constant. Consequently, the calculation
does not address the dynamics leading to the phenomenological coupling
constants. In contrast, the analysis presented in this paper considers
the triangle diagrams as the driving terms and the loop diagrams as
corrections. By solving the coupled equations, the $\rho BB'$ coupling
constants and ranges are obtained as the self-consistent solutions of
the model considered. Hence, these solutions contain valuable
information on the dynamical content of the coupling constants.

We used the following partial-wave decomposition for the
$\rho BB'$ vertex functions:
$$ F_{\rho a;c}(p_{\rho},p_{a};p_{c})=
 <I_{a}t_{a},1t_{\rho}\mid (I_{a}1)I_{c}t_{c}>
 \sum_{Sm_{S}Lm_{L}}  <J_{a}\nu_{a},1\nu_{\rho}\mid (J_{a}1)Sm_{S}>$$
\begin{equation}
 \;\;\;\;\;\;\;\;\;\;\;\;\;\;\;
\times <Sm_{S},Lm_{L}\mid (SL)J_{c}\nu_{c}>
 Y^{*}_{Lm_{L}}(\hat{p}){\cal F}_(p_{\rho},p_{a},p_{c})\ ,
\label{eq:2.0}
\end{equation}
with the radial vertex function ${\cal F}$ parametrized by
\begin{equation}
 {\cal F}(p_{\rho},p_{a},p_{c})= {{\scrig}}_{(L)\rho a;c}\frac{1}
{\sqrt{2w_{c}} }
{p}^{L} v_{L}(\Lambda_{\rho a;c},p)\ .
\label{eq:2}
\end{equation}
The $p_{i}\ (i=\rho, a, c)$ denote the four-momenta of the {\it i}$\;$th
particle, and ${\bf p}_{i}$ is its spatial part; $p$ denotes the
magnitude of the relative
momentum between $\rho$ and $a$, and $w_{c}$ is the invariant mass of $c$.
The two Clebsch-Gordan  coefficients specify, respectively, the isospin and
angular momentum coupling schemes with $I$ and $J$ denoting the isospin
and spin of the particles, and $t$, $\nu$ their z-components.
The vertex function ${\cal F}$ is a scalar and depends on two independent
four-momenta which can be chosen as $p_{\rho}+p_{a}$ and $p_{\rho}-p_{a}$.
It is convenient to work in the $\rho-a$ c.m. system where the external
four-momenta have the simple expressions
$p_{\rho}=(\rho^{0},{\bf p})$, $p_{a}=(a^{0},\  -{\bf p})$, and
$p_{c}=(w_{c},{\bf 0})$. Clearly, $w_{c}=a^{0}+ \rho^{0}$. From the
two independent four-momenta one can form three independent invariant
scalars. We will put the baryon $a$ on the mass shell so that
$a^{0}=E_{a}({\bf p})=\sqrt{m_{a}^2+{\bf p}^{2}}$. Consequently,
${\cal F}$ will depend only on two independent variables which we choose
as $w_{c}(\equiv w)$ and $p(\equiv\mid{\bf p}\mid)$. Because the
parametrization used in Eq.(\ref{eq:2}) is not the most general one,
we should expect ${\scrig}$ and $\Lambda$ to depend on $w$. This
$w$-dependence has important physical consequences and will be discussed
later. In the literature, vertex functions have often been made to depend
only on one variable. We emphasize that the one-variable dependence is
exact only when two of the three particles are on their mass shells.

Parity conservation limits the values of orbital angular momentum in
Eq.(\ref{eq:2.0}) to $L=1$ for $\rho N\rightarrow N$ and
$\rho N\rightarrow\Delta$, and to $L=1$ and 3 for $\rho\Delta\rightarrow N$
and $\rho\Delta\rightarrow\Delta$. As the Lagrangian models in the
literature consider only $L=1$, we shall solve the coupled equations in the
p-wave channel and omit, henceforth, the index $L$ of ${\scrig}$ and $v$.
Parity conservation also
limits the $\rho\rightarrow\pi\pi$, $\pi N\rightarrow N$,
$\pi N\rightarrow\Delta$, and $\pi\Delta\rightarrow N$ processes to
relative p-wave interactions alone. On the other hand,
$\pi\Delta\rightarrow\Delta$ can have both  p- and f-wave interactions.
Again, only the p-wave interactions will be retained in the calculations
in order to make a close connection to the Lagrangian models. The
$\rho\pi\pi$ vertex function is parametrized as
\begin{equation}
 F_{\rho;\pi\pi}(p_{\rho},p_{1},p_{2})=\sum_{t_{1}t_{2}}
C_{I}\sum_{m}\frac{{\scrig}_{\rho;\pi\pi}}{\sqrt{2M_{\rho}}}
\bar{h}(\Lambda_{\rho\pi\pi},k)Y^{*}_{1 m}(\hat{k})\sqrt{2\omega_{\pi}({\bf
p}_{1})
2\omega_{\pi}({\bf p}_{2})}\ ,
\label{eq:2.2}
\end{equation}
where
$C_{I}\equiv <1t_{1},1t_{2}\mid (11)1t_{\rho}>,
\bar{h}\equiv kr/(1 + k^{2}r^{2})^{2}$ with $r\equiv 1/\Lambda_{\rho\pi\pi}$,
and $k=\mid{\bf k}\mid$ is the $\pi\pi$ relative momentum. The $\pi_{1}ab$
vertex function is parametrized by
\begin{equation}
 F_{\pi_{1}a;b}(p_{1},p_{a};p_{b})=\sum_{I_{b}t_{b}J_{b}\nu_{b}m_{1}}
C_{I} C_{J} \frac{{\scrig}_{\pi a;b}}{\sqrt{2\omega_{\pi}({\bf p}_{1})}}
h(\Lambda_{\pi a;b},\kappa)Y^{*}_{1 m_{1}}(\hat{\kappa})\ ,
\label{eq:2.3}
\end{equation}
where $C_{I}\equiv <I_{a}t_{a},1t_{1}\mid (J_{a}1)I_{b}t_{b}>$,
$C_{J}\equiv <J_{a}\nu_{a},1m_{1}\mid (J_{a}1)J_{b}\nu_{b}>$,
${\bf \kappa}$ is the $\pi a$ relative momentum, $m_{1}$ is the z-component
of the corresponding orbital angular momentum, and $h$ is a dipole form
factor defined by $h\equiv (\kappa/{\mpi})[(\Lambda^2-{\mpi}^2)
/(\Lambda^2+{\bf \kappa}^2)]^{2} $. The vertex functions $F_{\pi_{2}b;c}$,
$F_{\rho b;b'}$, $F_{a;\pi b}$, and $F_{\pi b';c}$
are parametrized in terms of the corresponding momenta in a form
similar to Eq.(\ref{eq:2.3}). The triangle as well as the loop diagrams
depend on one four-momentum integration variable $q=(q^{0}, {\bf q})$.
By closing the contour of $q^{0}$ along a semi-circle in either the
upper or the lower half-plane, we can carry out the $q^{0}$ integration
analytically. Upon introducing Eqs.(\ref{eq:2})-(\ref{eq:2.3}) into
Eq.(\ref{eq:1}) and projecting out the ($L =1$) angular momentum
dependence, we obtain a set of four coupled equations for $v$.

One should note that the value of a coupling constant depends on the
parametrization convention of a theory. It is, therefore, useful to
relate the ${\scrig}$'s of the present partial-wave formalism to the
coupling constants of other works. To this end, we have used the S-matrix
convention of ref.\cite{Cele}. In relation to the Lagrangian model and
the parametrization employed in ref.\cite{Mach}, we obtain
${\scrig}_{\pi NN} =\sqrt{(3/2{\pi}^2)}f_{\pi NN}$  and
${\scrig}_{\pi N\Delta}= f_{\pi N\Delta}/\sqrt{6{\pi}^2}$. For the
$\pi\Delta\Delta$ vertex, ${\scrig}_{\pi\Delta\Delta}=
(5/2) f'_{\pi\Delta\Delta}/\sqrt{6{\pi}^2}$ where
$f'_{\pi\Delta\Delta}\equiv ({\mpi}/2M_{\Delta})g_{\pi\Delta\Delta}$
and $g_{\pi\Delta\Delta}$ is the axial-vector coupling constant
defined in ref.\cite{Arnd}. With respect to the parametrization in
ref.\cite{Lee}, ${\scrig}_{\rho\pi\pi}/\sqrt{2M_{\rho}}=g_{\rho\pi\pi}$
and ${\scrig}_{\pi\Delta\Delta}=(15/4) f_{\pi\Delta\Delta}/\sqrt{6{\pi}^2}$.
Hence, $f_{\pi\Delta\Delta}=(2/3) f'_{\pi\Delta\Delta}$.
In addition, $f_{\pi N\Delta}$ and $g_{\rho\pi\pi}$ can be directly
related to the experimental widths of $\Delta$ and $\rho$.

The ${\scrig}_{\rho BB'}$ can also be readily related to the $f$'s
and $g$'s defined in the Lagrangian formalism. Upon expressing the
matrix elements $<B'\mid {\cal H}_{\rho BB'}\mid \rho B>$ of the
Lagrangian model in a partial-wave representation and equating them
to Eq.(\ref{eq:2}), we obtain at ${\bf p}^{2}=0$
\begin{equation}
  ({\scrig}v)_{\rho NN}\ 4(\sqrt{2}-1)
\sqrt{3{\pi}^{2}M_{\rho}M_{N}}= (fu)_{\rho NN}+(gu)_{\rho NN}=
(1+\kappa_{\rho})(gu)_{\rho NN}\ ,
\label{eq:5}
\end{equation}
\begin{equation}
  ({\scrig}v)_{\rho N\Delta}\sqrt{\frac{3}{2}}(\sqrt{5}+1)
\sqrt{3{\pi}^{2}M_{\rho}^{3}/M_{\Delta}}= (fu)_{\rho N\Delta} ,
\label{eq:6}
\end{equation}
where ${\scrig}v$ has the dimension of inverse mass, and
$\kappa_{\rho}\equiv (f/g)_{\rho NN}$. The $u$ denotes the specific
form factor employed in the Lagrangian model, which varies from one
published work to another. Using these two equations, one can make a
straightforward comparison between the di-pion and the MEP results.
Our calculation gives the sum $(f+g)_{\rho NN}$. A knowledge of
$\kappa_{\rho}$ from the analysis of the nucleon electromagnetic
form factor can be used to determine $f$ and $g$ separately.

For the calculation, the values $g_{\rho\pi\pi}=0.6684 M_{\pi}^{-1/2}$,
$\Lambda_{\rho\pi\pi}=2.336$ fm$^{-1}$, $f_{\pi NN}=1 $, and
$\Lambda_{\pi NN}=\Lambda_{\pi N\Delta}=950$ MeV$ (\equiv \Lambda_{(2)})$
were used. The $\rho\pi\pi$ parameters fit the $\pi\pi$ phase
shifts.\cite{Lee}\ The value of the dipole range $\Lambda_{(2)}$ is taken
from ref.\cite{Kuma}. It corresponds to a monopole range $\Lambda_{(1)}=
0.644\times 950=612$\ MeV, which is nearly half of the MEP pion
monopole range of 1.2 GeV.\cite{Mach}\cite{Aren} For the other parameters,
we used the following sets: (a) $f_{\pi N\Delta}=1.69$ (ref.\cite{Mach}),\
$ f_{\pi\Delta\Delta}=0.8$ (ref.\cite{Lee}),\ $ \Lambda_{\pi\Delta\Delta}=
\Lambda_{(2)};$\  (b) $f_{\pi N\Delta}=2.19$,\ $f_{\pi\Delta\Delta}=1.03$
(or $f'_{\pi\Delta\Delta}=1.55$),\
$\Lambda_{\pi\Delta\Delta}=\Lambda_{(2)}$;\ and (c) $f_{\pi N\Delta}=2.19,\
f_{\pi\Delta\Delta}=1.03,\ \Lambda_{\pi\Delta\Delta}= 1.86\ $GeV
corresponding to the extreme situation of having an equivalent
monopole range $\Lambda_{(1),\pi\Delta\Delta}=1.2$ GeV. The criteria
leading to the above choices are as follows. In set (a) the ratio
$f_{\pi\Delta\Delta}/f_{\pi N\Delta} = \sqrt{2}/3$ satisfies a SU(6) quark
model result.\cite{Weis} Using the experimental $\Gamma_{\Delta}= 115$\ MeV
to calculate $f_{\pi N\Delta}$ and using $g_{\pi\Delta\Delta}^2/4{\pi}= 60$
(given by the isobar analyses\cite{Arnd}\cite{Dicu} of the
$\pi N\rightarrow \pi\pi N$ data), we obtain set (b). As these isobar
analyses do not give $\Lambda_{\pi\Delta\Delta}$, we used
$\Lambda_{\pi\Delta\Delta}= 950$ MeV in set (b) and 1.86 GeV in set (c)
in order to see the effects of this range parameter. We have also
evaluated contributions by intermediate $N^{*}$(1440) state
in the triangle diagrams, using $f_{\pi NN^{*}}=0.467$ and
$f_{\pi\Delta N^{*}}=1.63$ calculated from the median values of experimental
partial
widths $\Gamma_{N^{*}(\pi N)}$ and $\Gamma_{N^{*}(\pi\Delta)}$.
The full, coupled-equation results are presented in Table 1. The values
within parentheses are due to the inclusion of the $N^{*}$ contribution.
The $(f+g)_{\rho NN}$ and $f_{\rho N\Delta}$ have been evaluated,
respectively, at $w_{c}=$ 939 and 1232 MeV; they are found to be
higher than the corresponding Born results by 5\% and 14\%. For the
convenience of making comparisons, we have listed the values of $f$ and
$g$, as obtained from Eqs.(\ref{eq:5}) and (\ref{eq:6}). Further,
$g_{\rho NN}$ has been calculated from the solutions $(f+g)_{\rho NN}$
with $\kappa_{\rho}=f/g=3.7$ given by an analysis of the electromagnetic
form factor of the nucleon.

The MEP values of $g_{\rho NN}$ vary from 1.28 to 2.27\ (i.e.
$g^{2}/4{\pi}= 0.13 - 0.41$).  Those for $f_{\rho N\Delta}$ vary
from 4.91 to 7.81\ (or $f^{2}_{\rho N\Delta}/4{\pi}= 1.92 - 4.86$).
These lower and upper limits are, respectively, the results of
refs.\cite{Gari} and \cite{Mach}. An inspection of Table 1 shows
that with the inclusion of $N^{*}$,
sets (b) and (c) lead to a $g_{\rho NN}$ that is very close to
the MEP values. In fact, a good agreement has been obtained
when we used $f_{\pi NN^{*}}$
given by the experimental upper bound of
$\Gamma_{N^{*}(\pi N)}$.
While parameters of set (a) give a real part of $f_{\rho N\Delta}$
that is smaller than the lower limit of the MEP values by a factor of
$\sim 2.5$, the use of $f_{\pi N\Delta}$ and $f_{\pi\Delta\Delta}$
derived directly from the data (set (b)) brings the difference down to
within 40\%. Use of a large $\Lambda_{\pi\Delta\Delta}$ (set (c))
makes the real part of $f_{\pi N\Delta}$ agree with the MEP values.
However, in view of the deep inelastic scattering data on the upper
limit of $\Lambda_{\pi NN}$ and $\Lambda_{\pi N\Delta}$, the large
$\Lambda_{\pi\Delta\Delta}$ of set (c) may be questionable. Consequently,
we consider the results due to set (b) as being more realistic.
Our results also indicate that the contribution by $N^{*}$ is small. At this
point, a comment on the complex nature of $f_{\rho N\Delta}$ is in order.
As pointed out after Eq.(\ref{eq:2}), the coupling constants can be
energy-dependent. An important feature of the di-pion model is that
$f_{\pi N\Delta}$ becomes complex-valued at $w > M_{\pi}+M_{N}\ \equiv
w_{th}$. The $f_{\rho N\Delta}$ of Table 1 were calculated at $w=M_{\Delta}
> w_{th}$, where the $\pi N$ channel is open, thereby, giving rise to
Im$[f_{\rho N;\Delta}] < 0$. The energy dependence of the real and
imaginary parts of the $\rho N\Delta$ form factor at ${\bf p}$=0 are
shown, respectively, as the solid and dashed curves in fig.2, where we
have defined $H(w,p)\equiv ({\scrig}v)(w,p)$.

\vspace{0.1in}
\begin{center}
Table 1: The coupling constants and ranges (in MeV) given by the di-pion
model\\
for the $\rho NN$ vertex at $w=M_{N}$ and for the $\rho N\Delta$
vertex at $w=M_{\Delta}$.

\begin{tabular}{|c|c|c|c|c|c|} \hline \hline

Set & $(f+g)_{\rho NN}$  & $g_{\rho NN}$ &
$\Lambda_{\rho NN}$ & $f_{\rho N\Delta}$ &
$\Lambda_{\rho N\Delta}$ \\  \hline
 a & 4.39(4.90) & 0.934(1.04) & 435   & $1.89-2.09i(2.08-2.09i)$ & 495  \\
\hline
 b & 4.82(5.32) & 1.025(1.13) & 430   & $3.50-2.71i(3.69-2.71i)$ & 490  \\
\hline
 c & 4.86(5.36) & 1.033(1.14) & 430   & $5.25-2.70i(5.36-2.70i)$ & 490  \\
\hline \hline
\end{tabular}
\end{center}
\vspace{0.25in}

Table 1 indicates that the ranges of the calculated form factors are
in the region of 450 MeV, much smaller than the MEP values
of $1.2$ GeV.\cite{Mach}\cite{Aren} Here, we are considering the form factors
as functions of ${\bf p}^{2}$ and define
$\Lambda$ as the momentum at which
the magnitude of form factor is half of its value at ${\bf p}^{2}=0$.
Similar ranges have been obtained when we used the four-momentum
transfer $p_{\rho}^{2}$ as the variable.
We can, in fact, easily show
that a smaller range is a consequence of the composite nature of a vertex.

We have further examined the four-pion content of the vertex functions
by adding the $\rho\rightarrow \omega\pi$ mechanism into the triangle
diagrams. Using parameters of set (b) together with
$g_{\omega\rho\pi}M_{\omega}$=9.71 (ref.\cite{Durs}),
$g_{\omega NN}$=11.54 (ref.\cite{Mach}), and assuming the relation
$g_{\omega\Delta\Delta}/g_{\omega NN} =g_{\pi\Delta\Delta}/g_{\pi NN}$ in
our calculations, we have found that contributions from four-pion dynamics
are negligible because $M_{\omega}\gg {\mpi}$. Although diagrams having
more than three subvertices could in principle also contribute, they tend
to be unimportant because more are the particle propagators less is the
interaction probability. It is, however, worth  recalling that while both
the $L=1$ and 3 interactions can contribute to the $\pi\Delta\Delta$
process, only the $L=1$ case is considered in this and other works.
Although inclusion of the $L=3$ interaction can be easily implemented
in the present formalism, experimental information on this f-wave
coupling is sparse and uncertain.
If one can establish experimentally that the f-wave coupling constant is not
too small, then it could make a sizable contribution because
the f-wave vertex function is $\propto q^{3}$ and $q$ is an
integration variable that extends to very large values.

In summary, the di-pion model predicts $g_{\rho NN}$ that
agrees with the MEP values, but the predicted $f_{\rho N\Delta}$
is about 30\% smaller. In view of the fact that the $g_{\rho NN}$ and
$f_{\rho N\Delta}$ of this work are based on a microscopic model while
those of MEP were correlated fitting parameters, we regard the results
given by the two approaches as being compatible. The inclusion
of four-pion dynamics via the $\omega\pi$ doorway state does not
alter the results, indicating that the di-pion mechanism does represent
the leading mesonic contribution. However, experiments capable of
providing information on f-wave $\pi\Delta\Delta$
couplings can help determine if mesonic dynamics alone is sufficient
to account for all the strength of the phenomenological $\rho N\Delta$
coupling constants given by MEP.
The ranges of the calculated form factors are in the $450$\ MeV region,
much smaller than the corresponding MEP values, but in line with the
small $\pi NN$ and $\pi N\Delta$ ranges.\cite{Fran}\cite{Kuma}
As mentioned earlier, a small range also reflects the
composite nature of a vertex function.
Smaller ranges will give calculated
meson-exchange contributions that are different from those given by
large ranges and, hence, may lead to new understandings of the data.
Our analysis has  shown that
in the medium-energy
regime, there is no compelling need for introducing non-hadronic dynamics.
We stress, however, that MEP theory will become impractical
at very high energies because the exchange of many heavy
mesons has to be
included in order to produce the correct energy dependence of the
$NN$ cross section. Furthermore, the theory will be deficient because
the quark substructure of the hadrons
will start to manifest. We also stress that
while it is a good approximation to employ real-valued vertex functions in
analyzing $NN$ scattering
below the pion production threshold,\cite{Mach}\cite{Aren}\
the situation is  different
in meson production experiments where the energy $w_{th}$ can be surpassed.
Thus, using real-valued
vertex functions is questionable. We believe that
a nonvanishing imaginary part of the form factor can give
new interference effects in nuclear reaction calculations. This aspect of
the meson-exchange dynamics merits a systematic investigation in the future.

We would like to thank Sarah Novotny for her assistance with computer
graphics. One of us (LCL) would also like to thank Drs. C.M. Shakin and
M. Manley for enlightening discussions. This work was done under the
auspices of U.S. Department of Energy.


\pagebreak

\begin{figure}
\vspace{6.5in}
\caption{ Feynman diagrams for $\rho BB'$ vertex functions.}
\end{figure}
\pagebreak
\begin{figure}
\vspace{6.5in}
\caption{ Re$[H(w,0)]$(solid curve) and $-$Im$[H(w,0)]$(dashed curve)
      of the Born $\rho N\Delta$ vertex function versus
      $(w-M_{N})/M_{\pi}$\ obtained with parameters of set (b).}
\end{figure}

\end{document}